\documentclass[letterpaper,12pt]{article}   
\usepackage{osajnl2} 
\usepackage[draft]{hyperref} 
\usepackage[]{amsmath}

\begin{document}

\title{A mid-infrared broadband achromatic astronomical beam combiner for nulling interferometry}

\author{Hsien-kai Hsiao,$^{1,*}$ Kim A. Winick,$^1$ and John D. Monnier$^2$}
\address{$^1$Department of Electrical Engineering and Computer Science, University of Michigan, \\ 1301 Beal Avenue, Ann Arbor, MI 48109, USA}
\address{$^2$Department of Astronomy, University of Michigan, \\ 500 Church Street, Ann Arbor, MI 48105, USA}
\address{$^*$Corresponding author: hkhsiao@umich.edu}

\begin{abstract}
Integrated optic beam combiners offer many advantages over conventional bulk optic implementations for astronomical imaging. To date, integrated optic beam combiners have only been demonstrated at operating wavelengths below 4 $\mu$m. Operation in mid-infrared wavelength region, however, is highly desirable. In this paper, a theoretical design technique based on three coupled waveguides is developed to achieve fully achromatic, broadband, polarization-insensitive, lossless beam combining. This design may make it possible to achieve the very deep broadband nulls needed for exoplanet searching.
\end{abstract}

\ocis{110.3175, 350.1260, 130.3120, 120.6168, 130.3060.}

\maketitle

\section{Introduction}
\label{Intro}
The direct imaging and characterization of Earth-like planets, especially those capable of supporting life, is one of the outstanding goals of modern astrophysics and science in general. Currently, high-angular resolution astronomical imaging can be achieved interferometrically by combining the wavefronts from spatially separated telescopes~\cite{Labeyrie06}. Interferometric imaging and nulling require that the light from multiple apertures be combined. Although beam combining can be performed using bulk optics, integrated optic (IO) implementations offer a number of important advantages. These include spatial filtering, enhanced stability, on-chip fringe scanning, compactness and scalability. IO beam combiners for astronomical imaging were first proposed by Kern, Malbet, Schanen-Duport, and Benech in 1996~\cite{Kern96}. Using silicate-based glass IO devices, laboratory and on-sky stellar interferograms were demonstrated at astronomical H (1.5 $\mu$m -- 1.8 $\mu$m) and K (2.0 $\mu$m -- 2.4 $\mu$m) bands~\cite{BergerII,BergerIII,BergerIV,BergerV,BergerVI}. In many situations, especially those involving nulling and exoplanet search, operation in the infrared beyond 3 $\mu$m, where silicate-based glasses are not transparent, is required. On-sky interferometric measurements have been performed in the L band (3.0 $\mu$m -- 4.0 $\mu$m) using a guided-wave device, consisting of a two-beam fluoride glass fiber coupler, but this fiber-based technology is not easily scalable to multiple apertures~\cite{Mennesson99}. 

We have recently developed a prototype, single-mode, integrated optic, astronomical, beam combiner fabricated by titanium-indiffusion in an x-cut, lithium niobate (LiNbO$_3$) wafer~\cite{HKH09,HKHFiO,HKH2010}. The device operates in the 3.2 $\mu$m -- 3.8 $\mu$m spectral region, which lies in the L band, and has on-chip, electro-optically (EO) controlled, fringe scanning capabilities. Our results confirm that IO devices are well suited to perform the beam combining function for astronomical imaging. In the infrared wavelength region, a host star is normally a million times brighter than the planet orbiting it, which presents major difficulties when trying to image the planet directly. Nulling interferometry offers the possibility to overcome this problem by attenuating the stellar light, thus enhancing the visibility of the planet. Generally, achromatic phase shifting and broadband achromatic beam combining functions are required for deep nulling, with nulling depths of 10$^{-6}$ or better over the infrared spectrum. Although we can image the planet by nulling at a single wavelength, there are key reasons to use a broad wavelength band. First, there are several key biomarkers in the infrared spectrum from 6 $\mu$m to 18 $\mu$m. Second, the total integration time needed to detect a planet increases as the spectral bandwidth is reduced.

One key component for achieving a broadband deep null is the broadband achromatic beam combiner. Generally, there are two basic types of planar, integrated optic beam combiners: reversed-Y combiners and directional couplers. The operation of a reversed-Y combiner is achromatic by symmetry. A reversed-Y combiner, however, when used as part of a nulling interferometer, suffers an inevitable 3 dB loss of signal. On the other hand, directional coupler type beam combiners can capture the entire signal, and hence are theoretically lossless provided that both interferometric outputs are recorded and subtracted. Unfortunately, the operation of the directional coupler is chromatic, and this wavelength dependence prevents the device from achieving deep broadband nulls. Various approaches have been suggested to mitigate chromaticity of 2 by 2 couplers and achieve a wavelength flattened response. These approaches include (i) asymmetric directional couplers, where the two constituent waveguides differ in width but are uniform along the direction of propagation~\cite{Takagi92}, (ii) Mach-Zehnder type interferometers	that are wavelength-insensitive~\cite{Jinguji90}, and (iii) tapered velocity couplers, where the waveguide widths are varied along the propagation direction~\cite{Milton75}. The operation of the devices described by (i) and (ii) above depend on the interference between modes, while the tapered velocity mode coupler (iii) is meant to operate adiabatically, where there is little coupling between the local normal modes. The later device is also referred to as a mode-evolution coupler. Unfortunately these approaches do not yield an ultra-broadband achromatic response nor do they preserve the initial phase difference of the input beams in the interference term at the output. Lack of broadband chromaticity in either power splitting or phase can seriously degrade the deep nulls required for exoplanets search. In this paper, we propose and present a theoretical design of an achromatic, broadband, polarization-insensitive, mode-evolution, integrated optic, beam combiner suited for space-based nulling interferometry. The design is based on a system of three coupled waveguides along the lines of~\cite{Schneider00} and~\cite{Ishikawa07}. The proposed device can be realized using germanium-on-silicon or germanium air-bridge waveguide structures~\cite{HKH2010}.

In Section~\ref{LNM}, the theory of normal modes of a three coupled waveguide system is presented. A design of an achromatic, polarization-insensitive, mode evolution beam combiner is proposed in Section~\ref{ProposedDesign}. In Section~\ref{CMTofLNM}, the derivation of the coupled mode equations that describe the mode coupling between the local normal modes is given, and a condition for adiabatic operation is presented. In Section~\ref{ABCDesign}, a numerical design example of the proposed achromatic beam combiner, based on candidate technologies for waveguide fabrication in mid-infrared wavelength region, is presented and the device performance is numerically evaluated. The results of the paper are summarized in Section~\ref{ABCDis}.

\section{Normal Modes of Three Coupled Waveguides}
\label{LNM}
\begin{figure}
	\centering
		\includegraphics[width=.7\textwidth]{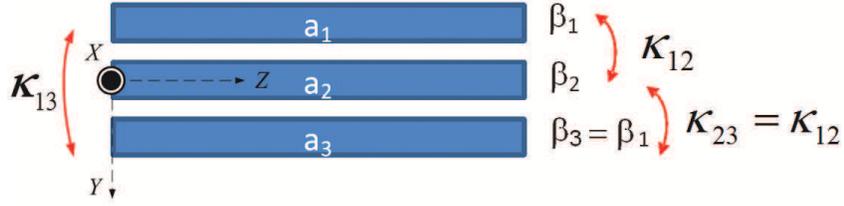}
	\caption{Three coupled channel waveguides. Waveguides 1 and 3 are identical and equidistant from waveguide 2.}
	\label{fig:ThreeWG}
\end{figure}

Before we go into the details of the design of the proposed fully-achromatic, broadband, mode-evolution beam combiner, we first develop representations for the normal modes of a symmetric three coupled waveguide system shown in the Fig.~\ref{fig:ThreeWG}~\cite{Tamir}. The waveguides shown in Fig.~\ref{fig:ThreeWG} are assumed to be uniform, i.e., do not change along the direction of propagation $z$. We also assume that (this assumption is valid when the waveguides are weakly coupled) the total scalar (TE or TM) field,$E(x,y,z)$, of the combined three waveguide structure can be approximately written as the weighted sum of the normalized waveguide modes, $\Phi_1(x,y)$, $\Phi_2(x,y)$, $\Phi_3(x,y)$, associated with each individual waveguide expressed in the single $x$, $y$,$z$ coordinate system shown in Fig.~\ref{fig:ThreeWG}. Thus,
\begin{align}
	E(x,y,z)=\Psi(x,y)\exp(j\beta z)=\sum_{l=1}^3a_l(z)\Phi_l(x,y) \label{Eq57a} \\
  \int_{-\infty}^{+\infty}\int_{-\infty}^{+\infty}\Phi_l^2(x,y)\;dxdy=1,\;\;\mbox{and $l$=1, 2, 3.} \label{Eq57b}	
\end{align}
where $\Psi(x,y)$ is a real-valued quantity which describes the spatial profile of the mode and $\beta$ is the associated propagation constant. The coupled-mode equations describing this system are~\cite{Tamir}
\begin{align}
	\frac{da_1}{dz}&=j\beta_1a_1+j\kappa_{12}a_2+j\kappa_{13}a_3 \nonumber \\		
	\frac{da_2}{dz}&=j\kappa_{12}a_1+j\beta_2a_2+j\kappa_{23}a_3  \nonumber \\
	\frac{da_3}{dz}&=j\kappa_{13}a_1+j\kappa_{23}a_2+j\beta_3a_3
	\label{Eq57}
\end{align}
where $\beta_1$ and $\beta_3$ are mode propagation constants for outer waveguides and $\beta_2$ is that of the center waveguide, $\kappa_{ij}$ is the coupling coefficient between waveguides $i$ and $j$, and $a_1$, $a_2$, and $a_3$ are the complex-valued mode amplitudes associated with the individual waveguide modes in waveguides 1, 2 and 3, respectively. We further assume that the outer waveguides are identical and equidistant from the center waveguide, and thus $\kappa_{12}$ = $\kappa_{23}$ and $\beta_1$ = $\beta_3$.

Considering a longitudinally varying waveguide structure, the amplitudes of the local normal modes are not constant as a function of position along the waveguide. These longitudinally-varying local normal modes will be designated by $\Psi_0(x,y;z)$, $\Psi_+(x,y;z)$ and $\Psi_-(x,y;z)$ and expressed as
\begin{align}
	E_l(x,y;z)&=\Psi_l(x,y;z)\exp(j\beta_l(z)z) \nonumber \\
						&=a_1^{(l)}(z)\Phi_1(x,y;z)+a_2^{(l)}(z)\Phi_2(x,y;z)+a_3^{(l)}(z)\Phi_3(x,y;z)
	\label{Eq62n} 
\end{align}
where $l=+, -, 0$ and $\Phi_n(x,y;z)$, $n=1, 2, 3$ are the local normal modes of the $n$ individual waveguides. Thus we define $A_l(z)$ as
\begin{align}
	A_l(z)=\exp(-j\beta_l(z)z)\left(\begin{array}{c}
					 	a_1^{(l)}\\a_2^{(l)}\\a_3^{(l)} \end{array}
					 \right),\;\;l=+, -, 0 
  \label{Eq62o}
\end{align}
and write
\begin{align}
	\Psi_l(x,y;z)=\left(\begin{array}{ccc}
					 			\Phi_1(x,y;z) & \Phi_2(x,y;z) & \Phi_3(x,y;z)
					 		\end{array}\right)A_l(z),\;\;l=+, -, 0
	\label{Eq62p}
\end{align}
By substituting Eq.~\eqref{Eq62p} into Eq.~\eqref{Eq57}, we obtain the following local normal modes of the three coupled waveguide structure
\begin{align}
	A_+=\left(\begin{array}{c}
					 	e/\sqrt{2}\\d\\e/\sqrt{2} \end{array}
					 \right),\,\,\,A_-=\left(\begin{array}{c}
					 	d/\sqrt{2}\\-e\\d/\sqrt{2} \end{array}
					 \right),\,\,\,A_0=\left(\begin{array}{c}
					 	1/\sqrt{2}\\0\\-1/\sqrt{2} \end{array}
					 \right) 
  \label{Eq62g}
\end{align}
and
\begin{equation}
	d=\sqrt{\frac{1}{2}(1+\frac{X}{\sqrt{1+X^2}})},\,\,\,e=\sqrt{\frac{1}{2}(1-\frac{X}{\sqrt{1+X^2}})}  
  \label{Eq62a}
\end{equation}
The waveguide parameter $X$ is defined as $X=\frac{\Delta\beta}{2\sqrt{2}\kappa_{12}}$, where $\Delta\beta=\beta_2-\beta_1-\kappa_{13}$. In general, the mode profile of a single local normal mode will evolve as it propagates along a longitudinally-varying waveguide structure. If we consider the three local normal modes of the combined longitudinally-varying three-waveguide structure, it is easily verified that $A_0(z)$, associated with the antisymmetric mode, remains constant along the whole structure. The other two symmetric local normal modes are identified by their mode order, i.e., by the relative magnitudes of their propagation constants, $\beta_+(z)$ and $\beta_-(z)$, respectively. The lowest order mode will be identified by the subscript $+$, and the third mode is identified by the subscript $-$ (where $\beta_+ > \beta_-$). Generally, power transfer occurs between the local normal modes of a longitudinally-varying structure. The degree of power transfer between the local normal modes is a function of the overlap parameter between their spatial mode profiles, $\Psi_i(x,y;z)$ and $\Psi_j(x,y;z)$ where $i,j\in\{-,0,+\}$. This overlap parameter is given by~\cite{Ishikawa07}
\begin{equation}
	<\Psi_i|\frac{\partial}{\partial z}|\Psi_j>\equiv\int_{-\infty}^{+\infty}\int_{-\infty}^{+\infty}\Psi_i^*(x,y;z)\frac{\partial}{\partial z}\Psi_j(x,y;z)\;dxdy
	\label{Eq62z}
\end{equation}
When $<\Psi_i|\frac{\partial}{\partial z}|\Psi_j>=0$, as occurs for a structure that does not change along the propagation direction $z$, there is no power exchange. Similarly if the structure varies slowly enough in z, then $\frac{\partial}{\partial z}\Psi_j(x,y;z)$ will be small and $<\Psi_i|\frac{\partial}{\partial z}|\Psi_j>$ will be approximately zero. Thus there will be little power exchange between the local normal modes. We propose a design of waveguide structures that are neither invariant along z nor vary slowly along z but for which $<\Psi_i|\frac{\partial}{\partial z}|\Psi_j>\approx 0$ and $i\neq j$. When there is very little power transfer between the local normal modes, the mode propagation is said to be \textit{adiabatic}.

\section{Proposed Design}
\label{ProposedDesign}
Based on the functional forms of the local normal modes of a three coupled waveguide system as shown in Eq.~\eqref{Eq62g}, we propose to use the symmetric three coupled waveguide structure shown in Fig.~\ref{fig:ABCLayout}(a) to achieve broadband achromatic beam combining. The waveguide parameters are given in Table~\ref{tab:WGParameters}. 

If monochromatic input beams with complex amplitudes $|a_1|\exp(j\phi_1)$ and $|a_3|\exp(j\phi_3)$, respectively, are launched into the outer waveguides 1 and 3, then the excited input field at $z$ = 0 can be expressed as the linear combination of the local normal modes $A_+(0)$ and $A_0(0)$ of the coupled structure at $z$ = 0 as indicated below
\begin{align}
	|a_1|\exp(j\phi_1)\left(\begin{array}{c}
	1 \\
	0 \\
	0
	\end{array}\right)&=|a_1|\exp(j\phi_1)[\frac{1}{2}\left(\begin{array}{c}
	1 \\
	0 \\
	-1
	\end{array}\right)+\frac{1}{2}\left(\begin{array}{c}
	1 \\
	0 \\
	1
	\end{array}\right)]	\nonumber \\
	&=|a_1|\exp(j\phi_1)[\frac{1}{\sqrt{2}}A_0(0)+\frac{1}{\sqrt{2}}A_+(0)]
	\label{Eq95}
\end{align}
\begin{align}
	|a_3|\exp(j\phi_3)\left(\begin{array}{c}
	0 \\
	0 \\
	1
	\end{array}\right)&=|a_3|\exp(j\phi_3)[\frac{-1}{2}\left(\begin{array}{c}
	1 \\
	0 \\
	-1
	\end{array}\right)+\frac{1}{2}\left(\begin{array}{c}
	1 \\
	0 \\
	1
	\end{array}\right)]	\nonumber \\
	&=|a_3|\exp(j\phi_3)[\frac{-1}{\sqrt{2}}A_0(0)+\frac{1}{\sqrt{2}}A_+(0)]
	\label{Eq95a}
\end{align}
\begin{align}
	\left(\begin{array}{c}
	|a_1|\exp(j\phi_1) \\
	0 \\
	|a_3|\exp(j\phi_3)
	\end{array}\right)&=[\frac{1}{\sqrt{2}}|a_1|\exp(j\phi_1)-\frac{1}{\sqrt{2}}|a_3|\exp(j\phi_3)]A_0(0) \nonumber \\
	&+[\frac{1}{\sqrt{2}}|a_1|\exp(j\phi_1)+\frac{1}{\sqrt{2}}|a_3|\exp(j\phi_3)]A_+(0)
	\label{Eq95b}
\end{align}
At the output of the device we wish to recover the input phase difference $\phi_1-\phi_3$ and the full power, i.e., $|a_1|^2+|a_3|^2$, of the incoming beams. The width of the center waveguide is kept fixed while the widths of the outer waveguides are equal to one another at all points along $z$ but are varied along $z$ to achieve the desired dephasing $\Delta\beta(z)=\beta_2(z)-\beta_1(z)-\kappa_{13}$. Note that the separation between waveguide 1 and 3 is assumed to be sufficiently large so that $\kappa_{13}\approx 0$ and thus can be neglected. The coupling term, $\kappa_{12}(z)$, and the dephasing term, $\Delta\beta(z)$, of the waveguide structure are chosen as indicated in Table~\ref{tab:WGParameters}. It will be shown in Sec.~\ref{ABCDesign} that the waveguide transition can be made nearly adiabatic (i.e., $<\Psi_i|\frac{\partial}{\partial z}|\Psi_j>\approx 0$, $i\neq j$). Thus the power in each local normal mode will remain constant as the mode propagates along the length of the structure, although the shape of each local normal mode will gradually evolve due to the adiabatic change of the waveguide parameters. The antisymmetric mode $A_0(0)$ and the lowest order symmetric mode $A_+(0)$ at $z=0$ will evolve into the following forms,
\begin{align}
	A_0(L)&=\left(\begin{array}{c}\frac{1}{\sqrt{2}}\\0\\\frac{-1}{\sqrt{2}}\end{array}\right)\exp(-j\int_0^L\beta_0(z)\;dz)
	\label{Eq102} \\
	A_+(L)&=\left(\begin{array}{c}0\\1\\0\end{array}\right)\exp(-j\int_0^L\beta_+(z)\;dz)
	\label{Eq103} 
\end{align}
respectively, at the output $z=L$. Thus by superposition the total output at $z=L$ becomes
\begin{align}	
	(|a_1|\exp(j\phi_1)-|a_3|\exp(j\phi_3))&\exp(-j\int_0^L\beta_0(z)\;dz)\frac{1}{2}\left(\begin{array}{c}1\\0\\-1\end{array}\right) \nonumber \\
	+& \nonumber \\
  (|a_1|\exp(j\phi_1)+|a_3|\exp(j\phi_3))&\exp(-j\int_0^L\beta_+(z)\;dz)\frac{1}{\sqrt{2}}\left(\begin{array}{c}0\\1\\0\end{array}\right)
	\label{Eq101}
\end{align}
Therefore, the intensities at outputs of waveguides 1, 2, and 3 at $z=L$ are given
\begin{align}
	\frac{1}{4}|\;|a_1|\exp(j\phi_1)-&|a_3|\exp(j\phi_3)\;|^2=\frac{1}{4}(|a_1|^2+|a_3|^2)-\frac{1}{2}|a_1||a_3|\cos(\phi_1-\phi_3) \nonumber \\
	&\mbox{: at output ports 1 and 3} \label{Eq104} \\
	\frac{1}{2}|\;|a_1|\exp(j\phi_1)+&|a_3|\exp(j\phi_3)\;|^2=\frac{1}{2}(|a_1|^2+|a_3|^2)+|a_1||a_3|\cos(\phi_1-\phi_3) \nonumber \\
	&\mbox{: at output port 2} \label{Eq105}
\end{align}
From these expressions we see that the total power in all three waveguides combined is $|a_1|^2+|a_3|^2$, and thus there is no loss in power. Furthermore, the input phase difference, $\phi_1-\phi_3$, is preserved in the output interference term, $|a_1||a_3|\cos(\phi_1-\phi_3)$, which is wavelength and polarization independent, guaranteeing fully achromatic and polarization-insensitive operation.

\section{Coupling of Local Normal Modes}
\label{CMTofLNM}
The definition of an adiabatic waveguide transition is a transition between two waveguide structures that takes place in such a manner that negligible power transfer occurs between the normal modes as they propagate from one structure to the other. Lack of mode conversion facilitates device design and can lead to wavelength insensitive operation as seen in Sec.~\ref{ProposedDesign}. Conventionally, the suppression of mode conversion has been achieved by slowly varying the waveguide structure along the direction of propagation. It was recently suggested, however, that a slow longitudinal variation, though sufficient, is not necessary to achieve adiabatic operation. More generally, the suppression of mode conversion is rooted in the small magnitude of the nonadiabatic term itself~\cite{Schneider00,Ishikawa07}. Thus conventional notions of adiabaticity based on slowly varying changes can be abandoned, and a new and different type of adiabaticity based on a controlled interaction can be exploited. In this section, the coupling equations governing the power transfer between local normal modes and the expression for the nonadiabatic term will be derived.

\subsection{Quasi-vector wave equation}
\label{WaveEq}
\begin{figure}
	\centering
		\includegraphics[width=.95\textwidth]{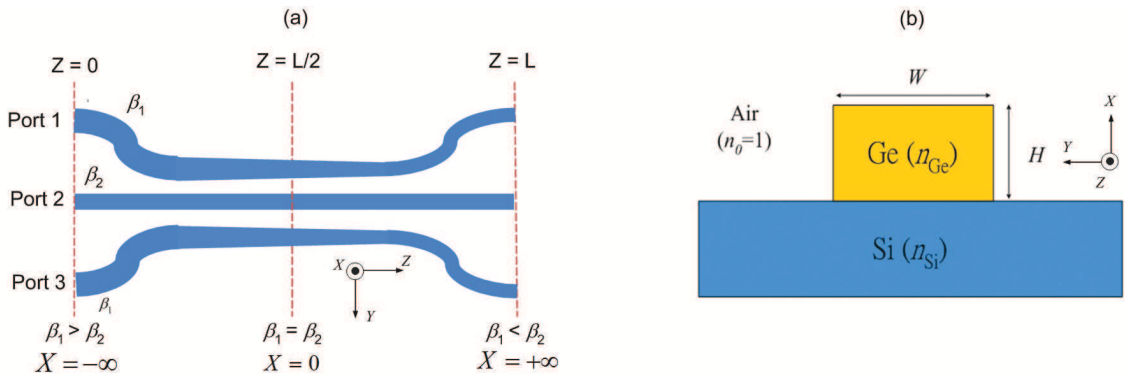}
	\caption{Schematic plot of proposed achromatic beam combiner.}
	\label{fig:ABCLayout}
\end{figure}

In order to calculate the spatial variation of electric field $\textbf{E}(x,y,z)$ and magnetic field $\textbf{H}(x,y,z)$ of the optical waveguide structure shown in Fig.~\ref{fig:ABCLayout}, we will use the vector wave equation derived from time-harmonic form of Maxwell's equations 
\begin{equation}
	(\nabla^2+k_0^2n^2)\textbf{E}+2\nabla((\nabla\ln n)\cdot \textbf{E})=0.
	\label{Eq67}
\end{equation}
The time dependence of the field is of the form $\exp(-j\omega t)$. The dielectric constant $\epsilon(x,y,z)$ of a waveguide is related to its refractive index $n(x,y,z)$ by $\epsilon=\epsilon_0n^2$, where $\epsilon_0$ is the free space electric permittivity, and the magnetic permeability is assumed to be its free space value ($\mu=\mu_0$) everywhere. $k_0=\omega/c=2\pi/\lambda_0$ is the free space wavenumber, $c=1/\sqrt{\mu\epsilon_0}$ is the vacuum speed of light and $\lambda_0$ is the wavelength of light in free space.

Although it is not necessary to do so, we will adopt a scalar analysis, and thus the last term in Eq.~\eqref{Eq67} will be neglected. Considering quasi-TE mode (i.e., $E_y \gg E_x$ and $E_y \gg E_z$) or quasi-TM mode (i.e., $E_x \gg E_y$ and $E_x \gg E_z$), the scalar (either y-component TE or x-component TM) electric field of the waveguide can be written as 
\begin{equation}
	E_i(x,y;z)=\Phi_i(x,y;z)\exp(j\beta_0^{(i)}z),\;\;\mbox{$i=x$ or $y$}
	\label{Eq68}
\end{equation}
where $\Phi_i(x,y;z)$ is the slowly varying field profile along the propagation direction $z$, $\beta_0=n_0k_0$ is the nominal propagation constant and $n_0$ is the nominal effective index of the optical mode. If the waveguides have no abrupt changes, then the transverse field $\Phi_i(x,y;z)$ will vary very little on the scale of a wavelength, and hence the paraxial approximation, $|\frac{\partial^2\Phi_i(x,y;z)}{\partial z^2}|\ll\beta_0|\frac{\partial \Phi_i(x,y;z)}{\partial z}|$ is valid. If we substitute Eq.~\eqref{Eq68} into Eq.~\eqref{Eq67}, apply the paraxial approximation, and replace the vector operator $\nabla^2$ by the scalar Laplacian $\nabla^2$, we obtain
\begin{align}
	-j\frac{\partial}{\partial z}\Phi_x = \frac{1}{2n_0k_0}&\left(\frac{\partial \Phi_x}{\partial x^2}+\frac{\partial \Phi_x}{\partial y^2}\right)+\frac{k_0(n^2(x,y;z)-n_0^2)}{2n_0}\Phi_x \mbox{  for TM;} \label{Eq68b} \\
	-j\frac{\partial}{\partial z}\Phi_y = \frac{1}{2n_0k_0}&\left(\frac{\partial \Phi_y}{\partial x^2}+\frac{\partial \Phi_y}{\partial y^2}\right)+\frac{k_0(n^2(x,y;z)-n_0^2)}{2n_0}\Phi_y \mbox{  for TE;}  
	\label{Eq68a}
\end{align}
Next if we substitute $\Phi_x=E_x\exp(-j\beta_0^{(x)}z)$ or $\Phi_y=E_y\exp(-j\beta_0^{(y)}z)$ back into Eq.~\eqref{Eq68b} or~\eqref{Eq68a}, we obtain the following equation
\begin{equation}
	-j\frac{\partial}{\partial z}E_i(x,y;z)=B_i(z)\cdot E_i(x,y;z),
	\label{Eq69}
\end{equation}
where $i=y$ for the quasi-TE mode and $i=x$ for the quasi-TM mode. The propagation operator $B_i(z)$ is given by
\begin{align}
	B_x(z) = \frac{1}{2n_0k_0}&\left(\frac{\partial}{\partial x^2}+\frac{\partial}{\partial y^2}\right)+\frac{k_0(n^2(x,y;z)+n_0^2)}{2n_0} \mbox{  for TM;} \label{Eq70aa} \\
	B_y(z) = \frac{1}{2n_0k_0}&\left(\frac{\partial}{\partial x^2}+\frac{\partial}{\partial y^2}\right)+\frac{k_0(n^2(x,y;z)+n_0^2)}{2n_0} \mbox{  for TE.}
  \label{Eq70}
\end{align}
It can be easily shown that the operator $B_i(z)$ is Hermitian. For the sake of simplicity in exposition, we will restrict the remaining discussion to the TE-mode and will designate the operator $B_y(z)$ simply as $B(z)$. The analysis for the TM-mode is similar. The three local normal modes $\Psi_0(x,y;z)$, $\Psi_+(x,y;z)$ and $\Psi_-(x,y;z)$ of the waveguide structure shown in Fig.~\ref{fig:ABCLayout} are eigenfunctions of Eq.~\eqref{Eq69}, i.e.,
\begin{equation}
	B(z)\Psi_l(x,y;z)=\beta_l(z)\Psi_l(x,y;z),\;\;l=+, -, 0.
	\label{Eq69a}
\end{equation}

\subsection{Coupling equation of local normal modes}
\label{CouplingEqOfLNM}
In the following analysis, the derivation of the coupling equation of the local normal modes will be carried out. Consider a system of three coupled waveguides with corresponding local normal modes $\Psi_0(x,y;z)$, $\Psi_+(x,y;z)$ and $\Psi_-(x,y;z)$, which are assumed to be real-valued (without loss of generality) since the operator $B(z)$ appearing in Eq.~\eqref{Eq69a} is Hermitian. The local normal modes will be normalized such that
\begin{align}
	<\Psi_0(x,y;z)|\Psi_0(x,y;z)>&=<\Psi_+(x,y;z)|\Psi_+(x,y;z)> \nonumber \\
															 &=<\Psi_-(x,y;z)|\Psi_-(x,y;z)> \nonumber \\
															 &=1,
	\label{Eq70a}
\end{align} 
and the inner product $<f(x,y;z)|g(x,y;z)>$ written in bra-ket notation is defined as
\begin{equation}
	<f(x,y;z)|g(x,y;z)>=\int_{-\infty}^{+\infty}\int_{-\infty}^{+\infty}f^*(x,y;z)g(x,y;z)\;dxdy.
	\label{Eq70b}
\end{equation}
It follows from Eq.~\eqref{Eq70a} and the fact that $\Psi_0(x,y;z)$, $\Psi_+(x,y;z)$ and $\Psi_-(x,y;z)$ are real-valued that
\begin{align}
	<\Psi_l(x,y;z)|\frac{\partial}{\partial z}|\Psi_l(x,y;z)>=0,\;\;l=0, +, -.
	\label{Eq70c}
\end{align}
Since the operator $B(z)$ is Hermitian it also follows that the local normal modes are orthogonal, that is
\begin{equation}
	<\Psi_l(x,y;z)|\Psi_k(x,y;z)>=0,\;\;l\neq k,\;\;\{l, k\}=0, +, -.
	\label{Eq70d}
\end{equation}
and therefore
\begin{align}
	<\Psi_l(x,y;z)|\frac{\partial}{\partial z}|\Psi_k(x,y;z)>&=-<\Psi_k(x,y;z)|\frac{\partial}{\partial z}|\Psi_l(x,y;z)>, \label{Eq70e} \\
																													 &l\neq k,\;\;\{l, k\}=0, +, - \nonumber 
\end{align}
Radiation modes will be neglected, and thus the total electric field, $E(x,y,z)$, in the coupled waveguides can be written as the following linear combination of the local normal modes
\begin{equation}
	E(x,y,z)=a_0(z)\Psi_0(x,y;z)+a_+(z)\Psi_+(x,y;z)+a_-(z)\Psi_-(x,y;z).
	\label{Eq70f}
\end{equation} 
Substituting the above equation into the paraxial wave equation~\eqref{Eq69}
\begin{equation}
	-j\frac{\partial}{\partial z}E(x,y,z) = B(z)E(x,y,z)
	\label{Eq70g}
\end{equation}
yields
\begin{align}
	&-j\frac{\partial a_0(z)}{\partial z}\Psi_0(x,y;z)-ja_0(z)\frac{\partial \Psi_0(z)}{\partial z} \nonumber \\
	&-j\frac{\partial a_+(z)}{\partial z}\Psi_+(x,y;z)-ja_+(z)\frac{\partial \Psi_+(z)}{\partial z} \nonumber \\
	&-j\frac{\partial a_-(z)}{\partial z}\Psi_-(x,y;z)-ja_-(z)\frac{\partial \Psi_-(z)}{\partial z} \nonumber \\
	&=a_0(z)B(z)\Psi_0(x,y;z)+a_+(z)B(z)\Psi_+(x,y;z)+a_-(z)B(z)\Psi_-(x,y;z) \nonumber \\
	&=a_0(z)\beta_0(z)\Psi_0(x,y;z)+a_+(z)\beta_+(z)\Psi_+(x,y;z)+a_-(z)\beta_-(z)\Psi_-(x,y;z)
	\label{Eq70h}
\end{align}
where $\beta_0(z)$, $\beta_+(z)$ and $\beta_-(z)$ are the local propagation constants associated with the local normal modes $\Psi_0(x,y;z)$, $\Psi_+(x,y;z)$ and $\Psi_-(x,y;z)$. Thus
\begin{align}
 B(z)\Psi_0(x,y;z)&=\beta_0(z)\Psi_0(x,y;z)
 \label{Eq70ii}\\
 B(z)\Psi_+(x,y;z)&=\beta_+(z)\Psi_+(x,y;z)
 \label{Eq70i}\\
 B(z)\Psi_-(x,y;z)&=\beta_-(z)\Psi_-(x,y;z).
 \label{Eq70j}
\end{align}
Multiplying both sides of Eq.~\eqref{Eq70h} by $\Psi_+(x,y;z)$, integrating over $x,y$ and using Eqs.~\eqref{Eq70a},~\eqref{Eq70c} and~\eqref{Eq70d}, yields
\begin{align}
	-j\frac{\partial a_+(z)}{\partial z}&-ja_0(z)<\Psi_+(x,y;z)|\frac{\partial}{\partial z}|\Psi_0(x,y;z)> \nonumber \\
	                                    &-ja_-(z)<\Psi_+(x,y;z)|\frac{\partial}{\partial z}|\Psi_-(x,y;z)>=a_+(z)\beta_+(z).
	\label{Eq70k}
\end{align} 
Similarly by multiplying both sides of Eq.~\eqref{Eq70h} by $\Psi_0(x,y;z)$ or $\Psi_-(x,y;z)$ and integrating over $x,y$ yields
\begin{align}
	-j\frac{\partial a_0(z)}{\partial z}&-ja_+(z)<\Psi_0(x,y;z)|\frac{\partial}{\partial z}|\Psi_+(x,y;z)> \nonumber \\
	                                    &-ja_-(z)<\Psi_0(x,y;z)|\frac{\partial}{\partial z}|\Psi_-(x,y;z)>=a_0(z)\beta_0(z).
	\label{Eq70kk}
\end{align}
\begin{align}
	-j\frac{\partial a_-(z)}{\partial z}&-ja_0(z)<\Psi_-(x,y;z)|\frac{\partial}{\partial z}|\Psi_0(x,y;z)> \nonumber \\
	                                    &-ja_+(z)<\Psi_-(x,y;z)|\frac{\partial}{\partial z}|\Psi_+(x,y;z)>=a_-(z)\beta_-(z).
	\label{Eq70l}
\end{align}
Equations~\eqref{Eq70k},~\eqref{Eq70kk} and~\eqref{Eq70l} can be written in matrix form as follows
\begin{equation}
	-j\frac{\partial}{\partial z}\left(
\begin{array}{c}
a_0(z)\\a_+(z)\\a_-(z)	
\end{array}\right)=\left(
\begin{array}{ccc}
\beta_0(z) & +j\xi_{0+}(z) & +j\xi_{0-}(z) \\
-j\xi_{0+}(z) & \beta_+(z) & +j\xi_{+-}(z)\\
-j\xi_{0-}(z) & -j\xi_{+-}(z) & \beta_-(z)	
\end{array}\right)\cdot\left(
\begin{array}{c}
a_0(z)\\a_+(z)\\a_-(z)	
\end{array}\right)
	\label{Eq70m}
\end{equation}
where
\begin{equation}
	\xi_{lk}(z)\equiv<\Psi_l(x,y;z)|\frac{\partial}{\partial z}|\Psi_k(x,y;z)>,\;\;l\neq k,\;\;\{l, k\}=0, +, -.
	\label{Eq70n}
\end{equation}
When there is no coupling (i.e., power exchange) between the local normal modes $\Psi_0(x,y;z)$, $\Psi_+(x,y;z)$ and $\Psi_-(x,y;z)$, the modes are said to evolve adiabatically. From Equation~\eqref{Eq70m}, it is clear that mode evolution will occur adiabatically when $\xi_{lk}(z)=0,\;\;l\neq k\;\;\{l, k\}=0, +, -$ for all $z$. It follows from Eq.~\eqref{Eq70n} that mode evolution will be adiabatic when the waveguide structure is uniform or changes slowly along the propagation direction $z$ since for these cases
\begin{equation}
	\frac{\partial}{\partial z}\Psi_l(x,y;z)=0\;\mbox{or}\;\approx 0,\;\;l=0, +, -.
	\label{Eq70p}
\end{equation}
Equation~\eqref{Eq70p}, however, need not be satisfied for adiabatic operation, and the more general condition is given by $\xi_{lk}(z)=0,\;\;l\neq k\;\;\{l, k\}=0, +, -$. In future discussions, $\xi_{lk}(z)$ will be referred to as the nonadiabatic terms. According to the results in Sec.~\ref{LNM}, the three local normal modes of the three waveguide structure can be written as
\begin{align}
 \Psi_+(x,y;z)&=a_1^{(+)}(z)\Phi_1(x,y;z)+a_2^{(+)}(z)\Phi_2(x,y;z)+a_1^{(+)}(z)\Phi_3(x,y;z)
 \nonumber \\
 \Psi_-(x,y;z)&=a_1^{(-)}(z)\Phi_1(x,y;z)+a_2^{(-)}(z)\Phi_2(x,y;z)+a_1^{(-)}(z)\Phi_3(x,y;z)
 \nonumber \\
 \Psi_0(x,y;z)&=a_1^{(0)}(z)\Phi_1(x,y;z)-a_1^{(0)}(z)\Phi_3(x,y;z)
 \label{Eq70q}
\end{align}
where $\Phi_l(x,y;z),\;\;l=1, 2, 3$ are the local normal modes of each of the three individual waveguides. Without loss of generality, the $\Phi_l(x,y;z),\;\;l=1, 2, 3$ will be assumed to be real-valued and normalized such that
\begin{equation}
	<\Phi_l(x,y;z)|\Phi_l(x,y;z)>=1,\;\;l=1, 2, 3.
	\label{Eq70r}
\end{equation}
It follows from Eq.~\eqref{Eq70r} that
\begin{equation}
	<\Phi_l(x,y;z)|\frac{\partial}{\partial z}|\Phi_l(x,y;z)>=0,\;\;l=1, 2, 3.
	\label{Eq70s}
\end{equation}
Since the two outer waveguides are identical and symmetrically placed relative to the central waveguide, it also follows that
\begin{align}
	<\Phi_1(x,y;z)|\frac{\partial}{\partial z}|\Phi_1(x,y;z)>&=<\Phi_3(x,y;z)|\frac{\partial}{\partial z}|\Phi_3(x,y;z)> \nonumber \\
	<\Phi_1(x,y;z)|\frac{\partial}{\partial z}|\Phi_3(x,y;z)>&=<\Phi_3(x,y;z)|\frac{\partial}{\partial z}|\Phi_1(x,y;z)> \nonumber \\
	<\Phi_2(x,y;z)|\Phi_1(x,y;z)>&=<\Phi_2(x,y;z)|\Phi_3(x,y;z)> 
	\label{Eq70t}
\end{align}
Combining Eqs.~\eqref{Eq70q}-~\eqref{Eq70t}, it may be easily verified that
\begin{align}
	\xi_{0+}(z)=<\Psi_0(x,y;z)|\frac{\partial}{\partial z}|\Psi_+(x,y;z)>=0 \nonumber \\
	\xi_{0-}(z)=<\Psi_0(x,y;z)|\frac{\partial}{\partial z}|\Psi_-(x,y;z)>=0.
	\label{Eq70u}
\end{align}
Consequently,
\begin{equation}
	-j\frac{\partial}{\partial z}\left(
\begin{array}{c}
a_0(z)\\a_+(z)\\a_-(z)	
\end{array}\right)=\left(
\begin{array}{ccc}
\beta_0(z) & 0 & 0 \\
0 & \beta_+(z) & +j\xi_{+-}(z)\\
0 & -j\xi_{+-}(z) & \beta_-(z)	
\end{array}\right)\cdot\left(
\begin{array}{c}
a_0(z)\\a_+(z)\\a_-(z)	
\end{array}\right).
	\label{Eq70v}
\end{equation}
Thus for adiabatic operation we need only consider the coupling between the symmetric local normal modes $\Psi_+(x,y;z)$ and $\Psi_-(x,y;z)$.

\subsection{Waveguide structure and nonadiabatic term}
\label{NonAdiabatic}
\begin{figure}
	\centering
		\includegraphics[width=.8\textwidth]{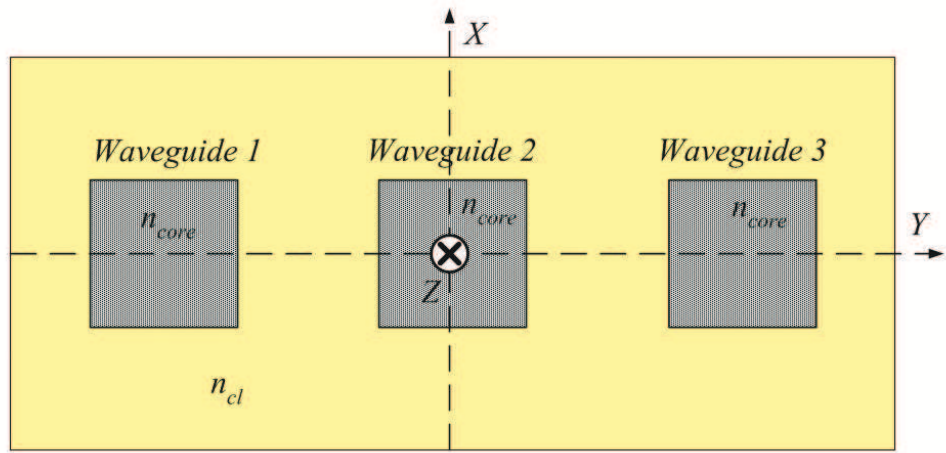}
	\caption{Cross-section view of the coupled two-waveguide system along with refractive index values.}
	\label{fig:WGProfile}
\end{figure}

Because the derivation of the nonadiabatic term, $\xi_{+-}(z)$, is similar for both quasi-TE and -TM modes, we now will only consider the quasi-TE mode and denote the corresponding operator $B(z)$ as $B_{1+2+3}(z)$, $B_{1+3}(z)$, and $B_2(z)$ for the coupled three-waveguide system, the coupled waveguide system consisting of the individual waveguides 1 and 3 alone, and the individual waveguide 2 alone, respectively, with the following distribution of dielectric constants:
\begin{align}
 n_{1+2+3}^2(x,y;z)=&n_{cl}^2+\Delta n_1^2(x,y;z)+\Delta n_2^2(x,y;z) \nonumber \\
                    &+\Delta n_3^2(x,y;z):\mbox{coupled three-waveguide}
 \label{Eq71}\\
 n_{1+3}^2(x,y;z)=&n_{cl}^2+\Delta n_1^2(x,y;z)+\Delta n_3^2(x,y;z):\mbox{waveguides 1 and 3}
 \label{Eq72}\\ 
 n_2^2(x,y;z)=&n_{cl}^2+\Delta n_2^2(x,y;z):\mbox{waveguide 2 alone}
 \label{Eq73}\\
 \Delta n_i^2(x,y;z)=&n_{core}^2(x,y;z)-n_{cl}^2:\mbox{inside cores $i=1,2,3$, or 0 otherwise,}
 \label{Eq74}
\end{align}
where $n_{core}$ and $n_{cl}$ are the refractive indices of the waveguide cores and cladding regions, respectively. The waveguides and refractive index profiles are shown in Fig.~\ref{fig:WGProfile} for a buried waveguide structure. The local normal modes of the coupled system and the constituent waveguide modes are defined as the eigenmodes of the corresponding $B(z)$ operator.
\begin{align}
 B_{1+2+3}(z)\cdot \Psi_q(x,y;z)&=\beta_q(z)\cdot \Psi_q(x,y;z)
 \label{Eq75} \\
 B_{1+3}(z)\cdot \Psi_1(x,y;z)&=\beta_{1+3}(z)\cdot \Psi_1(x,y;z)
 \label{Eq75a} \\
 B_2(z)\cdot \Psi_2(x,y;z)&=\beta_2(z)\cdot \Psi_2(x,y;z).
 \label{Eq76}
\end{align}
where the index $q=\{+,-\}$ is used to distinguish the lowest ($+$) and the other ($-$) symmetric local normal modes of the coupled three waveguides system ($\beta_+ > \beta_-$). $\Psi_1(x,y;z)$ is the lowest order local normal mode associated with the coupled waveguide system consisting of waveguides 1 and 3 alone. $\Psi_2(x,y;z)$ is the fundamental mode associated with waveguide 2 alone. Since the $B$ operators are Hermitian, the eigenfunctions of $B_{1+2+3}(z)$ at each fixed z (i.e. the local normal modes) are orthogonal and the corresponding eigenvalues are real-valued. Without loss of generality, the local normal modes are normalized as follows:
\begin{equation}
  \left(\begin{array}{cc}
        <\Psi_+(x,y;z)|\Psi_+(x,y;z)> & <\Psi_+(x,y;z)|\Psi_-(x,y;z)>\\
        <\Psi_-(x,y;z)|\Psi_+(x,y;z)> & <\Psi_-(x,y;z)|\Psi_-(x,y;z)>	
        \end{array}\right)=
  \left(\begin{array}{cc}
        1 & 0\\
        0 & 1	
        \end{array}\right)
	\label{Eq80}
\end{equation}
It is easily verified that $\Psi_1(x,y;z)$ and $\Psi_2(x,y;z)$ are given by
\begin{align}
 \Psi_1(x,y;z)&=\frac{1}{\sqrt{2}}[\Phi_1(x,y;z)+\Phi_3(x,y;z)]
 \label{Eq80a} \\
 \Psi_2(x,y;z)&=\Phi_2(x,y;z)
 \label{Eq80b}
\end{align}
where $\Phi_1(x,y;z)$, $\Phi_2(x,y;z)$ and $\Phi_3(x,y;z)$ are the fundamental waveguide modes of the individual three waveguides. For this to be true, we need to assume weak coupling between waveguides 1 and 3. $S(z)$ is the overlap integral between $\Psi_1(x,y;z)$ and $\Psi_2(x,y;z)$, i.e., $S(z)\equiv<\Psi_1(x,y;z)|\Psi_2(x,y;z)>$, and thus
\begin{align}
  &\left(\begin{array}{cc}
        <\Psi_1(x,y;z)|\Psi_1(x,y;z)> & <\Psi_1(x,y;z)|\Psi_2(x,y;z)>\\
        <\Psi_2(x,y;z)|\Psi_1(x,y;z)> & <\Psi_2(x,y;z)|\Psi_2(x,y;z)>	
        \end{array}\right) \nonumber \\
        =&\left(\begin{array}{cc}
        1 & S(z)\\
        S(z) & 1	
        \end{array}\right)
	\label{Eq81}
\end{align}
When the fundamental modes of the individual waveguides are known and the coupling between the individual waveguides is not very strong, the local normal modes of the coupled system can be approximated as a linear combination of the individual waveguide modes and thus
\begin{equation}
  \left(\begin{array}{cc}
        \Psi_+(x,y;z) \\ \Psi_-(x,y;z)
        \end{array}\right)^t=
  \left(\begin{array}{cc}
        \Psi_1(x,y;z) \\ \Psi_2(x,y;z)
        \end{array}\right)^t\cdot
 \left(\begin{array}{cc}
        c_{1+}(z) & c_{1-}(z)\\
        c_{2+}(z) & c_{2-}(z)
        \end{array}\right)
	\label{Eq82}
\end{equation}
Substituting Eq.~\eqref{Eq82} into the definition of the local normal modes of the coupled system, i.e., Eq.~\eqref{Eq75}, and using Eq.~\eqref{Eq81}, we obtain the following eigenvalue equation for the coefficients $c_{iq}(z)$:
\begin{align}
  \left(\begin{array}{cc}
        B_{11}(z) & B_{12}(z)\\
        B_{21}(z) & B_{22}(z)
        \end{array}\right)&\cdot
  \left(\begin{array}{cc}
        c_{1+}(z) & c_{1-}(z)\\
        c_{2+}(z) & c_{2-}(z)
        \end{array}\right)=     
 \left(\begin{array}{cc}
        1 & S(z)\\
        S(z) & 1
        \end{array}\right) \nonumber \\
        &\cdot
  \left(\begin{array}{cc}
        c_{1+}(z) & c_{1-}(z)\\
        c_{2+}(z) & c_{2-}(z)
        \end{array}\right)\cdot
 \left(\begin{array}{cc}
        \beta_+(z) & 0 \\
        0 & \beta_-(z)
        \end{array}\right)                
	\label{Eq83}
\end{align}
where the matrix representation of the $B(z)$ operator of the coupled system and the coupling coefficients between constituent waveguides are, respectively, defined as
\begin{align}
  \left(\begin{array}{cc}
        B_{11}(z) & B_{12}(z)\\
        B_{21}(z) & B_{22}(z)
        \end{array}\right)=&
  \left(\begin{array}{cc}
        1 & S(z)\\
        S(z) & 1	
        \end{array}\right)\cdot
 \left(\begin{array}{cc}
        \beta_1(z) & 0 \\
        0 & \beta_2(z)
        \end{array}\right) \nonumber \\
        &+\left(\begin{array}{cc}
        \kappa_{11}(z) & \kappa_{12}(z)\\
        \kappa_{21}(z) & \kappa_{22}(z)
        \end{array}\right)                
	\label{Eq84}
\end{align}
\begin{equation}
  \kappa_{ij}(z)=\frac{k_0}{2n_0}\frac{<\Psi_i|\Delta N(3-j)|\Psi_j>}{<\Psi_j|\Psi_j>}
	\label{Eq85}
\end{equation} 
\indent where 
\begin{align}
\Delta N(1)=\Delta n_1^2(x,y;z)+&\Delta n_3^2(x,y;z)
	\label{Eq85b}
\end{align}
\begin{align}
\Delta N(2)&=\Delta n_{2}^2(x,y;z)
	\label{Eq85a}
\end{align}
If we treat the coupled waveguide system consisting of the individual waveguides 1 and 3 alone (i.e., $\Psi_1$) as waveguide \textit{a} and the individual waveguide 2 (i.e., $\Psi_2$) as waveguide \textit{b}, then the coupling coefficient, $\kappa_{ij}$, defined by Eq.~\eqref{Eq85}, is equivalent to the coupling coefficient between waveguides \textit{a} and \textit{b} provided that there is no coupling between individual waveguides 1 and 3 of the three coupled waveguide system. By taking the inner product of Eq.~\eqref{Eq82} with $\Psi_q(z)(q=+,-)$ and making use of the normalization condition Eq.~\eqref{Eq80} and Eq.~\eqref{Eq81}, the following additional constraint on $c_{iq}(z)$ is obtained
\begin{equation}
  \left(\begin{array}{cc}
        c_{1+}(z) & c_{1-}(z)\\
        c_{2+}(z) & c_{2-}(z)
        \end{array}\right)^t
  \left(\begin{array}{cc}
        1 & S(z)\\
        S(z) & 1
        \end{array}\right)
  \left(\begin{array}{cc}
        c_{1+}(z) & c_{1-}(z)\\
        c_{2+}(z) & c_{2-}(z)
        \end{array}\right)= 
  \left(\begin{array}{cc}
        1 & 0 \\
        0 & 1
        \end{array}\right)
	\label{Eq86}
\end{equation}
The solution of the simultaneous equations~\eqref{Eq83} and~\eqref{Eq86} can be obtained using the method developed in~\cite{Haus87} and further expanded upon by Ishikawa~\cite{Ishikawa07}. Weak coupling will be assumed, so that $|S(z)| \ll 1$. It will also be assumed that $\left|\frac{\kappa_{12}-\kappa_{21}}{\kappa_{12}}\right|\ll 1$ and $\left|\frac{\beta_1-\beta_2}{\beta_2}\right|\ll 1$. If we introduce new parameters $\theta(z)$ and $\phi(z)$ as
\begin{align}
	\tan\theta(z)&=\frac{\kappa_{12}^{\prime}(z)}{\delta_{12}^{\prime}(z)}	
  \label{Eq87}\\
	\tan\phi(z)&=\frac{S(z)}{\sqrt{1-S(z)^2}}		               
	\label{Eq88}
\end{align}
\indent where
\begin{align}
	\delta_{12}^{\prime}(z)&=\frac{\beta_1+\kappa_{11}-(\beta_2+\kappa_{22})}{2\sqrt{1-S^2}}	
  \label{Eq89}\\
	\kappa_{12}^{\prime}(z)&=\frac{\kappa_{12}+\kappa_{21}}{2(1-S^2)}-\frac{S}{1-S^2}\frac{\kappa_{11}+\kappa_{22}}{2},		               
	\label{Eq90}
\end{align}  
the solution of Eqs.~\eqref{Eq83} and~\eqref{Eq86} can be expressed as
\begin{align}
  \left(\begin{array}{cc}
        c_{1+}(z) & c_{1-}(z)\\
        c_{2+}(z) & c_{2-}(z)
        \end{array}\right)=\frac{1}{\cos\phi(z)} 
  \left(\begin{array}{cc}
        \cos\frac{\theta(z)+\phi(z)}{2} & -\sin\frac{\theta(z)+\phi(z)}{2} \\
        \sin\frac{\theta(z)-\phi(z)}{2} & \cos\frac{\theta(z)-\phi(z)}{2}
        \end{array}\right)
	\label{Eq91}
\end{align}
\begin{equation}
	\beta_{\pm}(z)=\bar{\beta}(z)\pm\sqrt{\delta_{12}^{\prime}(z)^2+\kappa_{12}^{\prime}(z)^2}
	\label{Eq92}
\end{equation}
\indent where
\begin{equation}
	\bar{\beta}(z)=\frac{\beta_1+\kappa_{11}+\beta_2+\kappa_{22}}{2(1-S^2)}-\frac{S}{2(1-S^2)}(\kappa_{12}+\kappa_{21}+S\beta_1+S\beta_2).
	\label{Eq92a}
\end{equation}
Using Eqs.~\eqref{Eq82} and~\eqref{Eq91}, the nonadiabatic term, $\xi_{+-}(z)=<\Psi_+|\frac{\partial}{\partial z}|\Psi_->$, can be expressed as follows:
\begin{align}
  \xi_{+-}(z)=-\frac{1}{2}\frac{\partial\theta(z)}{\partial z}&-\frac{\cos\theta(z)}{2\cos\phi(z)}\frac{\partial\phi(z)}{\partial z} \nonumber \\
  &-\frac{1}{\cos\phi(z)}<\Psi_2(x,y;z)|\frac{\partial}{\partial z}|\Psi_1(x,y;z)>.
	\label{Eq94}
\end{align}  
The nonadiabatic term is now expressed as a derivative of local waveguide parameters, i.e., coupling coefficients and propagation constants. Our goal is to minimize the nonadiabatic term, and thus design a nearly adiabatic device. If the center waveguide is chosen to be uniform along its length, then
\begin{equation}
	\frac{\partial}{\partial z}\Psi_2(x,y;z)=0
	\label{Eq94a}
\end{equation}
and thus (since $<\Psi_2|\frac{\partial}{\partial z}|\Psi_1>=-<\Psi_1|\frac{\partial}{\partial z}|\Psi_2>$) the last term in Eq.~\eqref{Eq94} vanishes.

\section{Numerical Simulation of Achromatic Beam Combiners}
\label{ABCDesign}
In practice, a fully adiabatic transition cannot be achieved. Therefore, there will be some power exchange between the local normal modes $\Psi_+(x,y;z)$ and $\Psi_-(x,y;z)$ as they propagate down the structure. This power coupling, however, can be minimized by careful waveguide design. The design process will be illustrated in this section by considering a germanium/silicon raised strip three coupled waveguide system. Equation~\eqref{Eq94} will be used to estimate the magnitude of the nonadiabatic term at each $z$ for the design example. The power that couples from local normal mode $\Psi_+(x,y;z)$ to $\Psi_-(x,y;z)$, as the local normal mode propagates from the beginning (i.e., $z=0$) to the end (i.e., $z=L$) of the device, will be determined by solving Eq.~\eqref{Eq70v} numerically. 

\subsection{Candidates of waveguide fabrication in mid-infrared region}
\label{CadWGFab}
Both chalcogenide elements~\cite{Ho06,Hu07} and group IV elements, especially silicon and germanium~\cite{Soref06}, have been studied as candidate materials for mid-IR waveguides. There are several promising substrate materials and fabrication methods for realizing mid-infrared waveguide circuits. The most promising candidates include:
\begin{enumerate}
	\item chalcogenide rib waveguides on chalcogenide substrates
	\item chalcogenide rib waveguides on silicon substrates
	\item silicon or germanium rib waveguide membranes with an air bridge underneath the rib to serve as a lower cladding layer
	\item Ge/Si heterostructure rib waveguides with the Ge rib on top of a Si substrate
	\item hollow air core ARROW waveguides with a SiGe/Si antiresonant cladding
\end{enumerate}

\begin{figure}
	\centering
		\includegraphics[width=.8\textwidth]{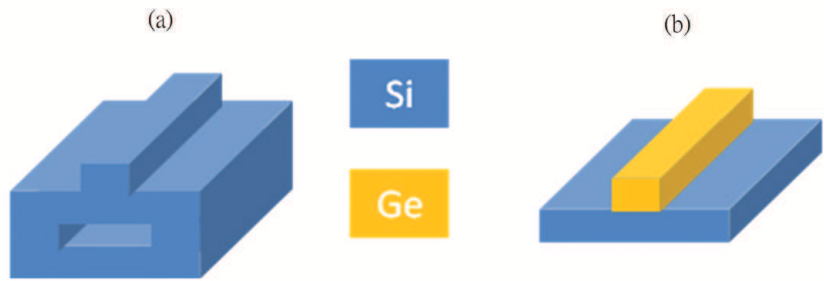}
	\caption{Proposed candidates for mid-infrared waveguide fabrication: (a) silicon (or germanium) rib membrane waveguide (b) Ge/Si heterostructure raised strip wave-guide.}
	\label{fig:MIRWG}
\end{figure}

The widespread use of silicon-based electronics, especially Si-CMOS technology, has become a major driving force beyond the development of silicon photonics~\cite{Soref2}. The successful developments of silicon photonic devices have aroused interests in deploying silicon for mid-infrared applications~\cite{Soref06}. Silicon has low loss in the spectral bands 1.2 $\mu$m -- 6 $\mu$m and 24 $\mu$m -- 100 $\mu$m, while multiphonon absorption prevents its use between 6 $\mu$m and 24 $\mu$m. EO modulation using the free carrier plasma effect becomes more efficient at wavelength beyond near IR, thus enhancing the functionality of silicon optical circuits in the mid-infrared regions. Traditional substrate cladding materials, such as SiO$_2$, which are suitable at near-IR are too lossy to be used for device operation beyond approximately 2 $\mu$m. Suspended rib waveguides, as shown in Fig.~\ref{fig:MIRWG}(a), consisting of silicon membrane clad below and above by air have been proposed as a potential solution~\cite{Soref06}. Another promising approach is to use a Ge/Si heterostructure consisting of a Ge raised strip waveguide of strain-relaxed crystal Ge epitaxially grown upon a silicon substrate as shown in Fig.~\ref{fig:MIRWG}(b)~\cite{Soref06}. Crystalline Ge has low loss from 1.9 $\mu$m out to approximately 12.5 $\mu$m. Under certain conditions, the high refractive index difference between Ge and Si may permit Si to be used as the lower cladding level even it is lossy in the 6 $\mu$m -- 12 $\mu$m band, since the penetration of the optical field into the cladding is limited. 

We will choose the Ge/Si heterostructure raised strip waveguide as the waveguide fabrication technique to illustrate our design method. The heteroepitaxial growth of multiple microns of high quality germanium film on silicon by multiple steps of growth and hydrogen annealing has been successfully demonstrated~\cite{Nayfeh06}. If the silicon cladding induces too much loss, a Ge rib waveguide with an air bridge underneath (see Fig.~\ref{fig:MIRWG}(a)) would be a logical choice instead.

\subsection{Numerical design and calculation results}
\label{NDofABC}
\begin{figure}
	\centering
		\includegraphics[width=.95\textwidth]{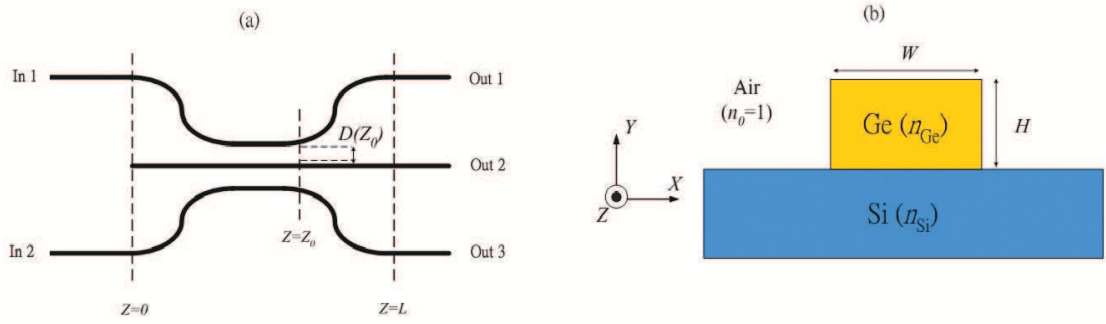}
	\caption{(a) Schematic of broadband achromatic beam combiner(b) Cross-section of Ge/Si raised strip waveguide geometry for fundamental mode calculation by beam propagation method.}
	\label{fig:GeSi2D}
\end{figure}

The schematic plot of the achromatic mode-evolution beam combiner is shown in Fig.~\ref{fig:GeSi2D} along with cross-section of the waveguide geometry based on Ge/Si raised strip waveguide. The refractive indices of the germanium strip and silicon substrate in the N band are taken from reference~\cite{HHLi} and are summarized in Table~\ref{tab:IndexOfGeSi}.

Due to relatively large refractive index difference between germanium and silicon, the optical mode of Ge/Si raised strip waveguide is well confined inside germanium strip. Thus, the results obtained for the buried waveguide as described in previous sections can be similarly applied to the waveguide geometry based on Ge/Si raised strip waveguide. Using the 3D structure together with a commercial, semi-vectorial 3D mode solver (RSoft BeamPROP), the waveguide parameters that correspond to single-mode operation in the N band at a nominal wavelength of $\lambda_0$ = 10 $\mu$m for both TE and TM modes are determined. The Ge strip height $H$ is chosen to be 3.5 $\mu$m and the strip width $W$ for the center waveguide is fixed to be 3.5 $\mu$m as well, while the widths of the outer waveguides shown in Fig.~\ref{fig:GeSi2D}(a) are varied from 3.58 $\mu$m at $z$ = 0 to 3.42 $\mu$m at $z$ = $L$. Since the central waveguide is uniform with width $W$ = 3.5 $\mu$m, the last term of Eq.~\eqref{Eq94} vanishes, and thus the nonadiabatic term is reduced to
\begin{align}
  \xi_{+-}(z)=-\frac{1}{2}\frac{\partial\theta(z)}{\partial z}-\frac{\cos\theta(z)}{2\cos\phi(z)}\frac{\partial\phi(z)}{\partial z}.
	\label{Eq94r}
\end{align}
Under weak coupling condition, $S(z)\approx 0$, and thus $\phi(z)\approx 0$ according to Eq.~\eqref{Eq88}. Therefore $\xi_{+-}(z)$ is approximately given by
\begin{align}
  \xi_{+-}(z)\approx-\frac{1}{2}\frac{\partial\theta(z)}{\partial z}.
	\label{Eq94rr}
\end{align}
Thus, nearly adiabatic operation of the device can be achieved by minimizing $\frac{\partial\theta(z)}{\partial z}$, where (see Eq.~\eqref{Eq87})
\begin{align}
  \tan\theta(z)=\frac{\kappa_{12}^{\prime}(z)}{\delta_{12}^{\prime}(z)}
	\label{Eq105a}
\end{align}
and (see Eqs.~\eqref{Eq89} and \eqref{Eq90})
\begin{align}
  \delta_{12}^{\prime}(z)&\approx\frac{\beta_1(z)+\kappa_{11}(z)-(\beta_2(z)+\kappa_{22}(z))}{2}
	\label{Eq105b} \\
	\kappa_{12}^{\prime}(z)&\approx\frac{\kappa_{12}(z)+\kappa_{21}(z)}{2}
	\label{Eq105c}
\end{align}
If the outer waveguides are identical to the center waveguide and uniform along the direction of propagation, it can be easily verified that
\begin{align}
  \kappa_{11}(z)&=\kappa_{22}(z)
	\label{Eq105dd} \\
	\kappa_{12}(z)&=\kappa_{21}(z)
	\label{Eq105ee}
\end{align}
according to Eqs.~\eqref{Eq85} - \eqref{Eq85a}. Thus if the variation of the width of the outer waveguides is much smaller than their nominal widths and the gap between the outer and the center waveguides, then the following relations approximately hold
\begin{align}
  \kappa_{11}(z)&\approx\kappa_{22}(z)
	\label{Eq105d} \\
	\kappa_{12}(z)&\approx\kappa_{21}(z)
	\label{Eq105e}
\end{align}   
Therefore, Eqs.~\eqref{Eq105a},~\eqref{Eq105b} and \eqref{Eq105c} can be approximately written as
\begin{align}
  \delta_{12}^{\prime}(z)&\approx\frac{\beta_1(z)-\beta_2(z)}{2}=\frac{\Delta\beta(z)}{2}
	\label{Eq105f} \\
	\kappa_{12}^{\prime}(z)&\approx\kappa_{12}(z)
	\label{Eq105g} \\
	\tan\theta(z)&\approx\frac{2\kappa_{12}(z)}{\Delta\beta(z)}
	\label{Eq105g2}
\end{align}
A three waveguide beam combiner is designed that achieves nearly adiabatic, achromatic operation for the TE mode at a center wavelength of $\lambda_0$ = 10 $\mu$m. The waveguide parameters (i.e., $\Delta\beta(z)$ and $\kappa_{12}(z)$) from $z=0$ to $z=L$ are varied according to Table~\ref{tab:WGParameters} such that $\frac{\partial\theta(z)}{\partial z}$ is small along the length of the device to achieve desired operation. The device's operation at 8 $\mu$m and 12 $\mu$m is also numerically evaluated. 
 
\begin{figure}[htbp]
  \centering
	\includegraphics[width=.9\textwidth]{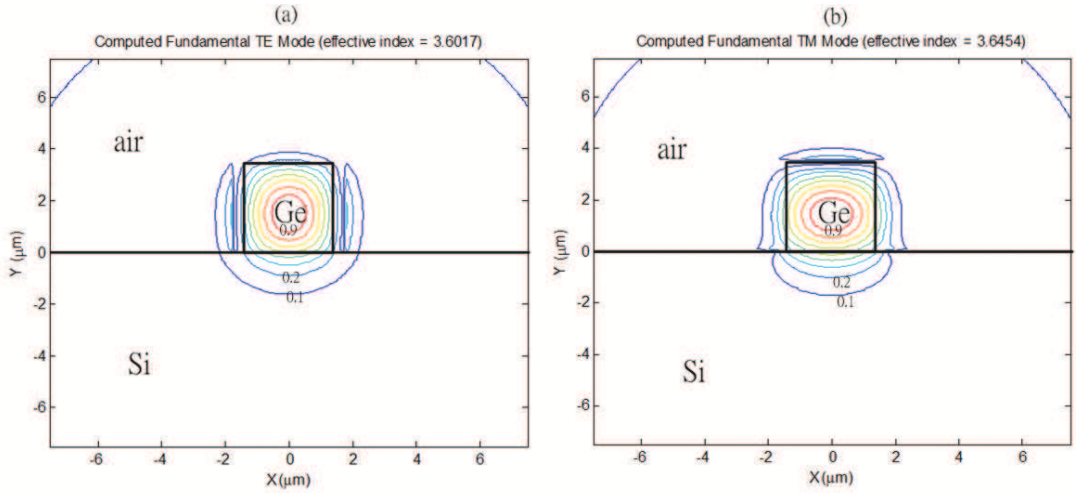}
	\caption{(a) TE mode profile at $\lambda_0$ = 10 $\mu$m, $n_{eff}$ = 3.6017 (b) TM mode profile at $\lambda_0$ = 10 $\mu$m, $n_{eff}$ = 3.6454 with nominal design ($H$ = $W$ = 3.5 $\mu$m).}
	\label{fig:GeSiTETMwl10}
\end{figure}

The fundamental TE and TM modes profiles for $H$ = $W$ = 3.5 $\mu$m, and $\lambda_0$ = 10 $\mu$m are shown in Fig.~\ref{fig:GeSiTETMwl10}. The effective refractive index, $n_{eff}$, of TE and TM modes are 3.6017 and 3.6454, respectively. The mode profiles are used to compute the coupling coefficient, $\kappa_{12}$, at different center-to-center gap spacings between outer and center waveguides for both TE and TM polarizations. The dephasing term, $\Delta\beta=\frac{2\pi}{\lambda_0}(n_{eff_2}-n_{eff_1})$, is also computed as a function of the width of the outer waveguides. Although not shown here, the coupling coefficients and the dephasing terms at 8 $\mu$m and 12 $\mu$m wavelengths have also been computed for the purpose of device verification at other wavelengths in the N band.

\begin{figure}
	\centering
		\includegraphics[width=.85\textwidth]{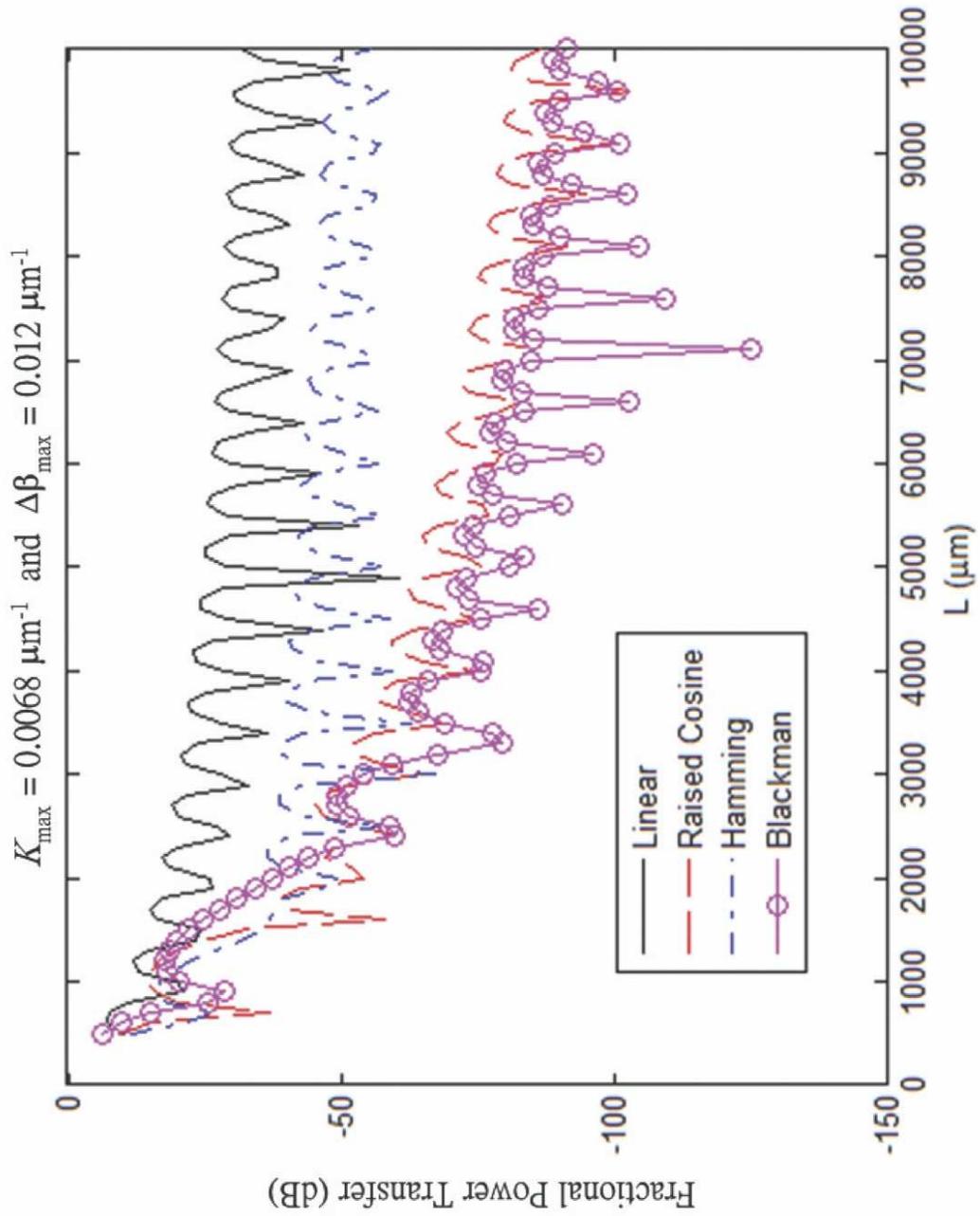}
	\caption{Power transferred from the TE local normal mode $\Psi_+(x,y;z)$ to the TE local normal mode $\Psi_-(x,y;z)$ at $\lambda_0$ = 10 $\mu$m.}
	\label{fig:ResiduePkVSL}
\end{figure}

\begin{figure}
	\centering
		\includegraphics[width=.75\textwidth]{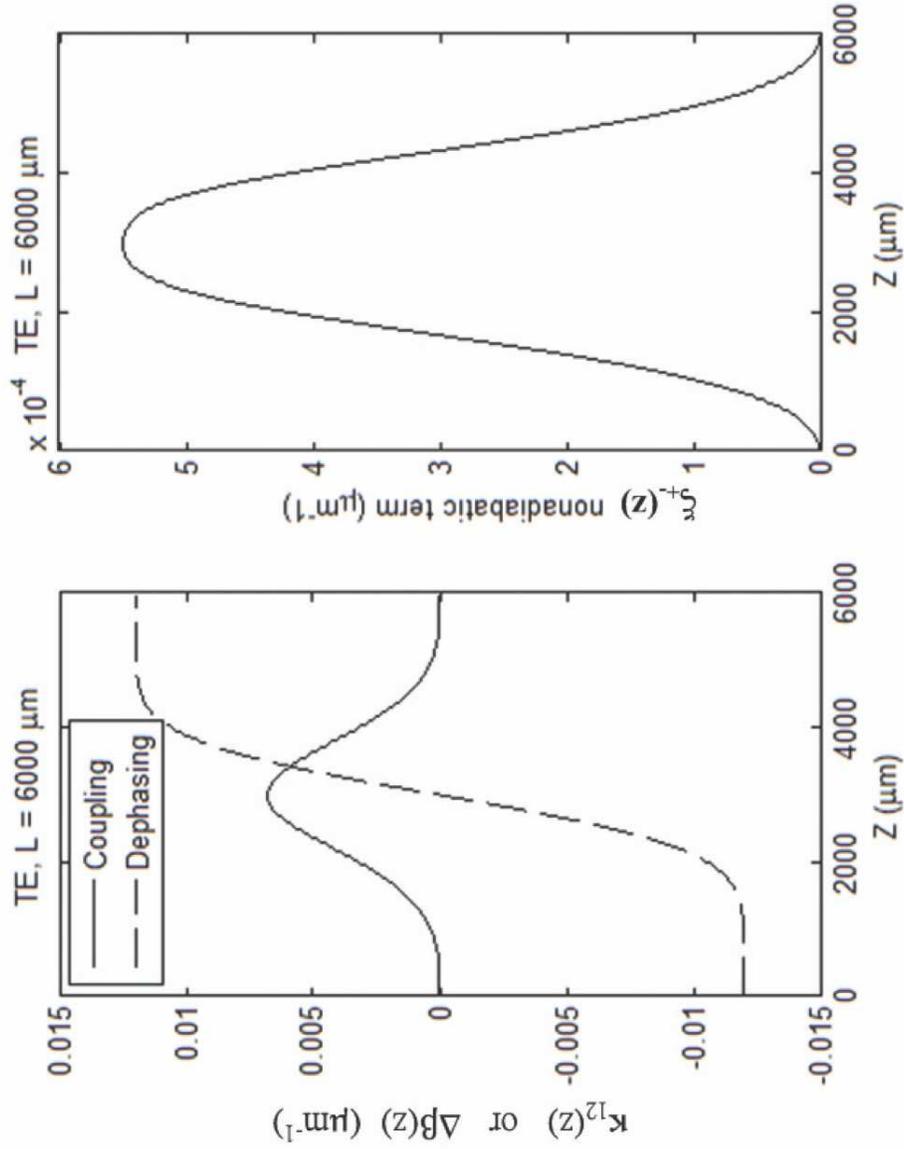}
	\caption{(Left) The variation of the coupling coefficient, $\kappa_{12}(z)$, and the dephasing term, $\Delta\beta(z)$, along propagation direction $z$ (Right) The nonadiabatic term, $\xi_{+-}(z)$, for the Blackman function for the TE mode at $\lambda_0$ = 10 $\mu$m.}
	\label{fig:KappaNonAdTerm}
\end{figure}

\begin{figure}
	\centering
		\includegraphics[width=.7\textwidth]{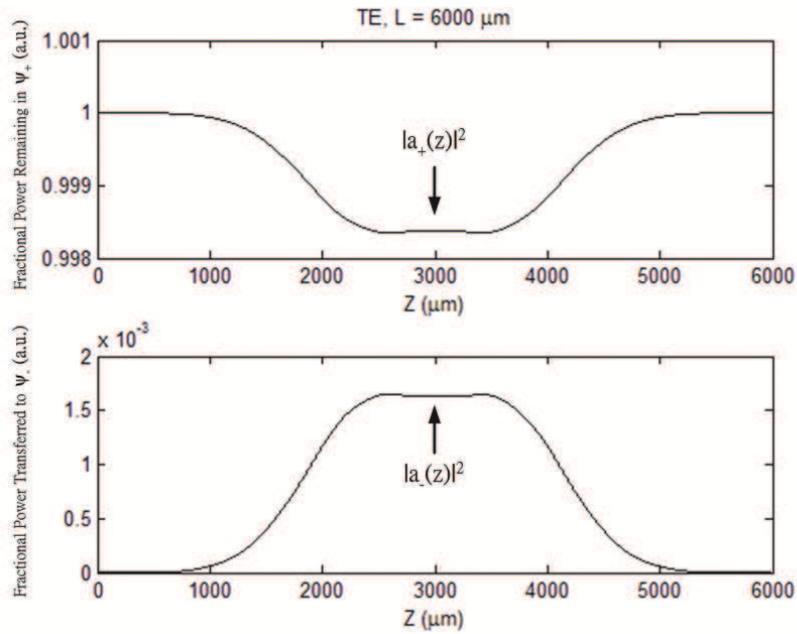}
	\caption{(Top) Total fraction of power remaining when only the local normal mode $\Psi_+$ is excited at $z=0$, i.e., $a_+(0)=1$ and $a_-(0)=0$. (Bottom) Fraction of launched power in $\Psi_+$ mode ($a_+(0)=1$ and $a_-(0)=0$) transferred to the local normal mode $\Psi_-$.}
	\label{fig:PowerTransferTE10}
\end{figure}

\begin{figure}
	\centering
		\includegraphics[width=.9\textwidth]{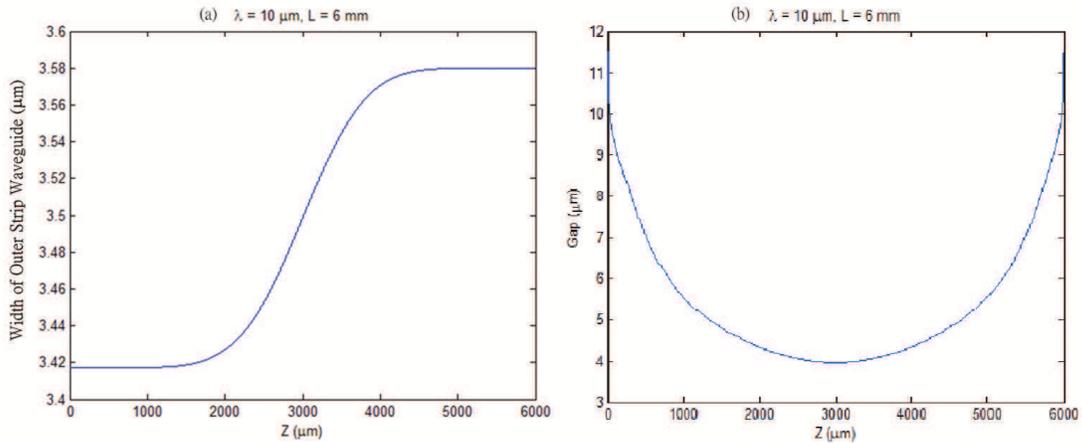}
	\caption{(a) The width variation of the outer waveguides as a function of propagation distance (b) The gap variation between the outer and the central waveguides as a function of propagation distance.}
	\label{fig:WidthDesign}
\end{figure}

The coupling coefficient $\kappa_{12}(z)$ between the outer waveguides and the center waveguide is restricted to positive values, while the dephasing term $\Delta\beta(z)=\beta_2-\beta_1$ can be either positive, 0 or negative. In order to estimate the power transferred from the $\Psi_+(x,y;z)$ to the $\Psi_-(x,y;z)$ local normal mode, we use the equivalent two-waveguide system described by Eq.~\eqref{Eq59}. 
\begin{align}
	\frac{da}{dz}&=j\beta_aa+jKb \nonumber \\		
	\frac{db}{dz}&=jKa+j\beta_bb
	\label{Eq59}
\end{align}
where
\begin{align}
	a(z)&=a_2(z)  \nonumber \\
	b(z)&=\sqrt{2}a_1(z)=\sqrt{2}a_3(z)
	\label{Eq106}
\end{align}
$a_1(z)$, $a_2(z)$, and $a_3(z)$ are the modal amplitude of each individual waveguide, and $a(z)$ and $b(z)$ are the equivalent modal amplitudes of the straight and curved waveguides as shown in Fig.~\ref{fig:GeSi2D}(a), respectively. The coupling coefficient $K(z)$ between the two waveguides in the two-waveguide system is related to the coupling coefficient $\kappa_{12}(z)$ between the outer and center waveguides of the three-waveguide system by $K(z)=\sqrt{2}\kappa_{12}$. As indicated earlier the design approach which will be taken is to vary $\Delta\beta(z)$ and $\kappa_{12}(z)$ along the length of the device in a manner consistent with Table~\ref{tab:WGParameters} and in such a way that $\frac{\partial\theta}{\partial z}\approx 0$, where $\tan\theta(z)\equiv\frac{2\kappa_{12}(z)}{\Delta\beta(z)}$. Such a design can be realized by choosing 
\begin{align}
	K(z)&=K_{max}\sin\vartheta(z)  
	\label{Eq107} \\
	\Delta\beta(z)&=-\Delta\beta_{max}\cos\vartheta(z)
	\label{Eq108}
\end{align}
where $K_{max}$ is the maximum coupling coefficient (which occurs at the device center $z=L/2$), $\Delta\beta_{max}$ is the maximum dephasing term (which occurs at the beginning and the end of the device, i.e., at $z=0$ and $L$), and $\vartheta(z)$ is a monotonically (or nearly monotonically) increasing function of $z$, with $\vartheta(0)=0$ and $\vartheta(L)=\pi$, that controls the rate of change of $\Delta\beta(z)$ and $\kappa_{12}(z)$. The waveguide parameter $X(z)$ is equal to $\frac{\Delta\beta(z)}{2K(z)}$. The nonadiabatic term according to Eq.~\eqref{Eq94rr} can be written as
\begin{equation}
  \xi_{+-}(z)\approx -\frac{1}{2}\frac{\partial\theta(z)}{\partial z}
	\label{Eq109}
\end{equation}
where $\tan\theta(z)\approx 1/X(z)$. Therefore, the nonadiabatic term can be expressed in terms of the variation of the waveguide parameter, $X(z)$, as follows
\begin{equation}
  \xi_{+-}(z)\approx \frac{1}{2(1+X^2)}\frac{\partial X}{\partial z}
	\label{Eq110}
\end{equation}
The coupling between the local normal modes $\Psi_+(z)$ and $\Psi_-(z)$ for the TE mode at $\lambda_0$ = 10 $\mu$m is computed with the nonadiabatic term described in Eq.~\eqref{Eq110} for the different $\vartheta(z)$ functions listed below~\cite{Schneider00}:
\begin{align}
	\vartheta(z)&=\frac{\pi z}{L}\;\mbox{: Linear Function}  
	\label{Eq111} \\
	\vartheta(z)&=\frac{\pi z}{L}-0.5\sin\frac{2\pi z}{L}\;\mbox{: Raised Cosine Function}  
	\label{Eq112} \\
	\vartheta(z)&=\frac{\pi z}{L}-0.426\sin\frac{2\pi z}{L}\;\mbox{: Hamming Function}  
	\label{Eq113} \\	
	\vartheta(z)&=\frac{\pi z}{L}-0.5952\sin\frac{2\pi z}{L}+0.0476\sin\frac{4\pi z}{L}\;\mbox{: Blackman Function}  
	\label{Eq114}
\end{align}
The maximum coupling coefficient, $K_{max}$, and the maximum dephasing term, $\Delta\beta_{max}$, are chosen to be 0.0068 $\mu$m$^{-1}$ and 0.012 $\mu$m$^{-1}$, respectively. These values are selected such that the single mode condition for each individual waveguide and the weak coupling condition between outer and center waveguides are maintained at wavelengths from 8 $\mu$m to 12 $\mu$m, as the waveguide parameters from $z=0$ to $z=L$ are varied according to Table~\ref{tab:WGParameters}. The fraction of power transferred (i.e., $-10\log_{10}|a_-(L)|^2$) from the $\Psi_+(x,y;z)$ local normal mode to the $\Psi_-(x,y;z)$ local normal mode is numerically computed using Eq.~\eqref{Eq70v} with $a_+(0)=1$ and $a_-(0)=0$. The result is shown for $L$ between 0 and 10000 $\mu$m for the various $\vartheta(z)$ functions and plotted in Fig.~\ref{fig:ResiduePkVSL}. Among different $\vartheta(z)$ functions, for $L >$ 3000 $\mu$m, the Raised Cosine and the Blackman functions outperform (i.e., smaller fractional power transferred from $\Psi_+(x,y;z)$ to $\Psi_-(x,y;z)$) the others and the Blackman function is slightly better than the Raised Cosine function. Based on Fig.~\ref{fig:ResiduePkVSL}, we choose $L$ = 6000 $\mu$m for our device. Better performance can always be achieved by using a longer value of $L$. For the Blackman function and $L$ = 6000 $\mu$m, the coupling term, $\kappa_{12}(z)$, and the dephasing term, $\Delta\beta(z)$, along with the corresponding nonadiabatic term, $\xi_{+-}(z)$, are shown in Fig.~\ref{fig:KappaNonAdTerm}, and the power transfer characteristics between the local normal modes along the propagation direction $z$ is shown in Fig.~\ref{fig:PowerTransferTE10}. Using computed coupling coefficients for both polarizations, together with Eqs.~\eqref{Eq107}, \eqref{Eq108}, and \eqref{Eq114} with $K_{max}$ = 0.0068 $\mu$m$^{-1}$, $\Delta\beta_{max}$ = 0.012 $\mu$m$^{-1}$, $L$ = 6000 $\mu$m, and $\lambda_0$ = 10 $\mu$m, the device layout is obtained. The width variation of the outer waveguides and the separation between the outer and center waveguides, as a function of the position $z$ along the device, are plotted in Fig.~\ref{fig:WidthDesign}(a) and~\ref{fig:WidthDesign}(b), respectively.

The performance of the broadband operation of the device design is also numerically evaluated using Eq.~\eqref{Eq70v} at wavelengths of 8 $\mu$m and 12 $\mu$m and the results are summarized in Table~\ref{tab:ResidueCheck}. Based on the data, it is seen that there is little coupling between the $\Psi_+$ and the $\Psi_-$ local normal modes, and hence the device operates adiabatically. Consequently, according to Sec.~\ref{ProposedDesign}, the performance of our mode-evolution beam combiner is relatively insensitive to the operating wavelength and the polarization, even though the coupling coefficient, $\kappa_{12}(z)$, and the dephasing term, $\Delta\beta(z)$, are highly wavelength-dependent.

The sensitivity of the performance of the device due to fabrication errors has also been numerically investigated. Cases where small perturbations to the coupling coefficient, $\kappa_{23}(z)$, or the propagation constant, $\beta_3(z)$, introduce asymmetries, i.e., $\kappa_{23}(z)\ne\kappa_{12}(z)$ and $\beta_3(z)\ne\beta_1(z)$, are considered. These perturbations are modeled here by the relationships
\begin{align}
	\delta=\frac{\kappa_{23}(z)-\kappa_{12}(z)}{\kappa_{12}(z)} \mbox{ \;\;or\;\; } \delta=\frac{\beta_{3}(z)-\beta_{1}(z)}{\beta_{1}(z)},
	\label{Eq116}
\end{align}
where the deviation $\delta$ is a constant independent of z. After introducing these perturbations, the coupled mode equations~\eqref{Eq57} are solved for the output (i.e., $a_1(L)$, $a_2(L)$, and $a_3(L)$) assuming that the symmetric mode $A_+(0)=(\frac{1}{\sqrt{2}}\; 0 \;\frac{1}{\sqrt{2}})^T$ (i.e., $a_1(0)=\frac{1}{\sqrt{2}}$, $a_2(0)=0$, and $a_3(0)=\frac{1}{\sqrt{2}}$) was launched into the three waveguide structure. If the device operates adiabatically then the output should all remain in the $A_+$ mode, i.e., $a_1(L)=0$, $a_2(L)=1$, and $a_3(L)=0$. The fraction of power coupled to the other modes, i.e., $1-a_2^2(L)$, is given in Table~\ref{tab:ResidueCheckDev} as a function of the deviation $\delta$. According to Table~\ref{tab:ResidueCheckDev}, the device operation is more sensitive to perturbations in the propagation constant than to perturbations in the coupling coefficient. The predictions in Table~\ref{tab:ResidueCheckDev}. however, may be a bit pessimistic in practice since they assume that the errors are constant along the entire length of the device.

\section{Conclusion}
\label{ABCDis}
We have presented the numerical design of a broadband, polarization insensitive, achromatic beam combiner for the operation in the astronomical N band based on Ge/Si heterostructure raised strip waveguides. The beam combining is intrinsically achromatic because of the symmetric arrangement of the three coupled waveguides. As opposed to a reversed-Y junction combiner which suffers a 3 dB loss, our device is theoretically lossless. Furthermore on-chip EO modulation is also possible by utilizing the free carrier plasma dispersion effect~\cite{Soref4} in silicon-based waveguides. Most importantly, the technology needed to actually fabricate the proposed design is quite promising and plausible. We believe the realization of such beam combiner will increase the possibility of using integrated optic combiners for space-based deep nulling interferometry.

\section*{Acknowledgments}
The authors gratefully acknowledge support from the National Aeronautics and Space Administration.



\clearpage
\begin{table}[htbp]
	\caption{Device parameters are varied adiabatically along the structure from $z=0$ to $z=L$.}
	\label{tab:WGParameters}	
	\begin{center}
		\begin{tabular}{cccccccccccc}
		\hline 
		\rule[-1ex]{0pt}{3.5ex}  Position $z$ && Parameter $X$ &&& $\Delta\beta$ &&& $\kappa_{12}$ &&& $\kappa_{13}$ \\ \hline 
    \rule[-1ex]{0pt}{3.5ex}  0    		&& $-\infty$   &&& $< 0$ &&& $\approx 0$ &&& $\approx 0$ \\ 
    \rule[-1ex]{0pt}{3.5ex}  $ L/2$   && 0   		&&& 0 	    				 &&& $\kappa_{max}$ &&& $\approx 0$			\\ 
		\rule[-1ex]{0pt}{3.5ex}	 $L$			&& $+\infty$ &&& $> 0$ &&& $\approx 0$  &&& $\approx 0$			\\  \hline
		\end{tabular}
	\end{center}		
\end{table}

\clearpage

\begin{table}[htbp]
	\centering
	\caption{Dispersion of Si and Ge at 20$^o$C and different mid-infrared wavelengths.}
		\begin{tabular}{ccccccc}
				\hline 
		           Temperature    && $\lambda (\mu m)$ 	&& $n$(Si)	 && $n$(Ge)  \\ \hline
      								   			&& 8 	&& 3.4158 && 4.0048  \\ 
      						20$^o$C		  && 10	&& 3.415 &&	4.0025  	\\
															&& 12	&& 3.4145 && 4.0012 	\\  \hline 							
		\end{tabular}
	
	\label{tab:IndexOfGeSi}
\end{table}

\clearpage  

\begin{table}[htbp]
	\centering
	\caption{Fractional power coupled into the TE and TM $\Psi_-$ mode at $L$ = 6000 $\mu$m evaluated at different wavelengths.}
		\begin{tabular}{ccccc}
				\hline 
		           	Wavelength				 &&			 $\Psi_-$ TE Mode    &&  $\Psi_-$ TM Mode \\ \hline
      					8 $\mu$m && -34 dB 	&& -24.1 dB    \\ 
      					10 $\mu$m && -82.2 dB	&& -39.5 dB   	\\
								12 $\mu$m  && -33.4 dB	&& -40.6 dB	\\  \hline 
		\end{tabular}
	\label{tab:ResidueCheck}
\end{table}
\clearpage

\begin{table}[htbp]
	\centering
	\caption{Fractional power coupled to other mode, i.e., $1-a_2^2(L)$, due to imperfection of the frabrication process.}
		\begin{tabular}{ccccc}
				\hline 
		           	Deviation in $\kappa_{23}(z)$				 &&			 Deviation in $\beta_3(z)$   &&  $1-a_2^2(L)$ \\ \hline
      					0.5 \% && 0 \% 	&& -52.0 dB    \\ 
      					1.0 \% && 0 \%	&& -46.1 dB   	\\
								5.0 \% && 0 \%	&& -32.3 dB	\\ 
								0 \% && 0.1 \%	&& -38.2 dB	\\
								0 \% && 0.5 \%	&& -24.3 dB	\\  
								0 \% && 1.0 \%	&& -18.3 dB	\\  	\hline
								 
		\end{tabular}
	\label{tab:ResidueCheckDev}
\end{table}

\clearpage

\section*{List of Figure Captions}
Fig. 1. Three coupled channel waveguides. Waveguides 1 and 3 are identical and equidistant from waveguide 2.

\noindent Fig. 2. Schematic plot of proposed achromatic beam combiner.

\noindent Fig. 3. Cross-section view of the coupled two-waveguide system along with refractive index values.

\noindent Fig. 4. Proposed candidates for mid-infrared waveguide fabrication: (a) silicon (or germanium) rib membrane waveguide (b) Ge/Si heterostructure raised strip wave-guide.

\noindent Fig. 5. (a) Schematic of broadband achromatic beam combiner(b) Cross-section of Ge/Si raised strip waveguide geometry for fundamental mode calculation by beam propagation method.

\noindent Fig. 6. (a) TE mode profile at $\lambda_0$ = 10 $\mu$m, $n_{eff}$ = 3.6017 (b) TM mode profile at $\lambda_0$ = 10 $\mu$m, $n_{eff}$ = 3.6454 with nominal design ($H$ = $W$ = 3.5 $\mu$m).

\noindent Fig. 7. Power transferred from the TE local normal mode $\Psi_+(x,y;z)$ to the TE local normal mode $\Psi_-(x,y;z)$ at $\lambda_0$ = 10 $\mu$m.

\noindent Fig. 8. (Left) The variation of the coupling coefficient, $\kappa_{12}(z)$, and the dephasing term, $\Delta\beta(z)$, along propagation direction $z$ (Right) The nonadiabatic term, $\xi_{+-}(z)$, for the Blackman function for the TE mode at $\lambda_0$ = 10 $\mu$m.

\noindent Fig. 9. (Top) Total fraction of power remaining when only the local normal mode $\Psi_+$ is excited at $z=0$, i.e., $a_+(0)=1$ and $a_-(0)=0$. (Bottom) Fraction of launched power in $\Psi_+$ mode ($a_+(0)=1$ and $a_-(0)=0$) transferred to the local normal mode $\Psi_-$.

\noindent Fig. 10. (a) The width variation of the outer waveguides as a function of propagation distance (b) The gap variation between the outer and the central waveguides as a function of propagation distance.

\clearpage

\end{document}